# Optics-less beam control of EUV high harmonics generated from solids

Byunggi Kim[1], Seungman Choi[1], Yong Woo Kim, Seung Jai Won, Young-Jin Kim and Seung-Woo Kim[*]

*Department of Mechanical Engineering, Korea Advanced Institute of Science and Technology (KAIST), 291 Daehak-ro, Yuseong-gu, Daejeon 34141, South Korea.*

[1]These authors equally contributed to this work.

*Corresponding author (swk@kaist.ac.kr)

**Abstract**

**High harmonics generation (HHG) of coherent extreme ultraviolet (EUV) radiation enables ultrafast pump-probe spectroscopy with attosecond-scale timing resolutions. Yet beam control of generated EUV radiation remains a major challenge for handy implementation of ultrafast spectroscopy particularly for nano-micro samples. Here, we demonstrate a solid-based HHG scheme that incorporates non-collinear illumination to perform the propagation control of EUV harmonics by wavefront phase manipulation via a spatial light modulator. Further, our solid-based non-collinear HHG scheme converts EUV harmonics to Bessel beams without extra optics for beam focusing, offering a useful tool for diverse EUV applications of metrology and fabrication.**

High harmonics generation (HHG) permits multiplying the driving laser frequency by nonlinear photon-electron interaction as well demonstrated by irradiating intense laser pulses on not only noble gases[1-3] but also bare solids[4,5], engineered solid surfaces[6,7], and liquid curtains[8]. Yet the optical frequency up-conversion by HHG has been exercised for diverse applications, with particular attention being paid to achieving extreme ultraviolet (EUV) radiation in the form of coherent ultrashort pulses for ultrafast observation of physical phenomena via pump-probe spectroscopy[9] or lens-less coherent diffraction imaging[10-12], even with angle-resolved photoemission spectroscopy[13]. Nonetheless, for more efficient implementation of ultrafast imaging and/or spectroscopic observation, generated EUV radiation needs to be well steered on targeted samples with strict control of the spatiotemporal coherence with respect to the driving laser pulse.

In general, EUV radiation is relatively not easy to handle because of its short wavelengths and, more severely, its attribute of being strongly absorbed in most optical materials. Routinely, for EUV beam delivery and focusing, complex reflective optics configured with a large incidence angle or structured with multi-layered metal films are widely adopted despite fabrication difficulty and cost. Even so, several handy schemes have been proposed to provide the capability of efficient EUV beam manipulation particularly without use of bulky reflective optics. First, Mimura et al.[14] demonstrated a method of wavefront correction for hard X-ray focusing with a sub-10-nm spot size using adaptive optics. Next, refractive optics was engineered using a gas jet[15], while EUV high harmonics was focused on a spot size of few microns by applying a spatial chirp to the driving laser field[16]. Another straightforward approach entailed the use of diffractive Fresnel zone plates for ptychographic diffraction imaging[17] and EUV ablation[18]. Further, Sivis et al.[6] directly generated harmonic beams from a zone-plate-patterned solid surface. Such diffractive focusing techniques proved to be successful in reducing the spot size to a few wavelengths, but still required extra EUV optics or filters to separate EUV harmonics from the driving laser field at a cost of system complexity and power loss.

Herein, we investigate a solid-based scheme of HHG devised to implement non-collinear illumination so as to separate EUV harmonics from the driving laser by the principle of momentum and parity conservation. In addition, non-collinear illumination leads EUV harmonics to be focused near the rear surface of the solid target, which is subsequently steered by wavefront manipulation of the driving laser field via a spatial light modulator. Further, our non-collinear HHG scheme is extended to a 3-dimensional configuration to convert EUV harmonics to Bessel beams without extra optics for beam focusing to a spot size of few wavelengths. Finally experimental results are discussed in detail with regards to the practical possibility to provide an efficient means of HHG for ultrafast pump-probe spectroscopy and fabrication using coherent EUV radiation.

**Results**

***Non-collinear high harmonics generation in solids.*** Figure 1a illustrates the experimental setup configured in this investigation to generate EUV high harmonics from ia bulk magnesium oxide (MgO) crystal. Non-collinear irradiation was realized by splitting the driving laser into two beams, of which the spatiotemporal propagation was aligned to be precisely symmetrical through a common-optics arrangement using a Wollaston prism, a Glan polarizer, a half-wave plate, a triangular optical wedge and a collimating lens (see Methods for details). We used a spatial light modulator (SLM) to control the propagation direction of the driving laser by inducing a linear phase variation across the wavefront. As a result, the target crystal is illuminated by two non-collinear laser beams symmetrically with an incident angle $\theta_d$ with respect to the surface normal, while . the $q$-th harmonic generated from the crystal leaves with a divergence angle $\theta_q$.

The divergence angle $\theta_q$ varies with the harmonic order $q$; Fig. 1b shows an experimental result for H7 – H17 obtained from a 150-μm thick MgO (100) crystal. In non-collinear irradiation using two driving lasers of k1 and k2 wave, the $q$th-order harmonic absorbs photons from two driving lasers, i.e. $q = q_1 + q_2$ with $q_1$ and $q_2$ being the integer are the numbers of photons absorbed from each driving pulse. the is determined in accordance to the momentum and parity conservation of the driving infrared (IR) photons. In photon models[19], from the driving IR laser beams in accordance with the parity conservation, As illustrated in Fig. 1b, the directionality of the harmonic beam with the smallest divergence angle well agrees with the sum of the driving wave vectors, i.e. $\tan\theta_q = \tan\theta_d/q$, where $\theta_q$ and $\theta_d$ are the divergence angles of $q$th harmonic and driving fields, respectively[20-22]. Figures 1c and d indicate satisfactory agreement between the intensity profiles of the harmonic beams and the parity conservation conditions. The multiple-harmonic beam evolution can also be interpreted with respect to the wave model, in which the far-field intensity is the Fourier transform of the driving field intensity at the sample, which corresponds to where the interference of the two driving beams acts as an active grating[20,21].

The phase-matching condition can be expressed by the wave vector description[20], i.e. $\Delta \mathbf{k} = q_1 \mathbf{k}_1 \cos(\theta_d - \theta_q) + q_2 \mathbf{k}_2 \cos(\theta_d + \theta_q) - \mathbf{k}_q$, where $\mathbf{k}_q$, $\mathbf{k}_1$, and $\mathbf{k}_2$ are the wave vectors of the $q$th harmonic and the first and second driving fields, respectively. In the case of a sparse medium, such as a noble gas, the refractive index changes due to plasma fraction strongly affects $\mathbf{k}_1$ and $\mathbf{k}_2$. Thus, the gas pressure and driving intensity were optimised to satisfy the phase-matching condition[22] $\Delta \mathbf{k} = 0$. Phase-matched HHG using a hollow-core waveguide[23] is a powerful source for a coherent EUV beam. On the other hand, the phase matching that occurs within the solid medium is negligible because of strong reabsorption of the EUV photons. Thus, it is reasonable to assume that the EUV emitters exist along a single layer of the surface. Accordingly, the high-harmonic field $E_q$ can be defined as

$$E_q(\mathbf{r},t) = q_{eff}(E_{s,surface})\exp[-j(\omega_q t + \mathbf{k}_q \cdot \mathbf{r})] \qquad (1)$$

where $\mathbf{r}$, $t$, $E_{s,surface}$, $j$ and $\omega_q$ denote the coordinate vector, time, driving field at the surface, imaginary unit, and high-harmonic angular frequency, respectively. The effective nonlinear coefficient $q_{eff}$ represents the phase and conversion efficiency of each harmonic corresponding to a driving field. In this study, we only considered conversion efficiency, and neglected temporal phase change of the high-harmonic field because multi-cycle driving pulses are used without stabilizing carrier envelope phase. As shown in Fig. 1e, the divergence angle of the harmonic beams was confirmed to be nearly independent of the driving intensity, which suggests that the nonlinear refractive index only minimally impacts direction of the harmonic wave vector $\mathbf{k}_q$.

Figure 2a shows the harmonic spectra obtained by the driving laser pulses with identical polarizations. The peak intensity of the driving laser pulse was fixed at 23 TW/cm$^2$. The measured high-harmonic spectra were found to exhibit four-fold symmetry owing to the face-centred cubic structure of the MgO crystal. Figure 2b shows that the yield of the higher-order harmonics (especially H11 and H15) increases in the Γ-X direction. This finding is supported by previous reports on HHG from a MgO crystal[24], which explained that dipole-allowed coupling is strong among the multiple conduction bands and the valence band in the Γ-X direction. Figure 2c shows the spectral yield for different driving intensities in the Γ-X direction. In the case of solid HHG[6,25], the transition from the perturbative scaling $I^q$ is commonly observed in coincidence with increasing driving intensity under a damage threshold of ~30 TW/cm$^2$. Thus, the results demonstrate that the electronic response originates from a crystal orientation in the NC geometry, whereas the driving field at the solid surface determines the directionality of the high-harmonic beams.

*Optics-less wavefront control of converging EUV beam*

Because the EUV wave vectors are determined according to the parity and momentum conservation within the nanometres skin layer, the EUV wavefronts can be manipulated by actively controlling the driving laser wavefronts and modifying the solid surface. We obtained a converging wavefront of EUV harmonics by overlapping two incident IR laser beams at a point 7 mm before their foci. Figures 3a and b show the beam profile of the 7th harmonic beam that was focused in the direction of propagation, as measured using the knife-edge method (see **Methods** for details). Owing to the NC geometry, the EUV beam could be measured without causing damage to any of the optical systems because it is spatially separated from the driving beam. The beam profile indicates that the EUV light sheet tends to form as the beam is focused in the *x* direction (horizontal) and collimated in the *y* direction (vertical). The femtosecond pulse duration of the driving laser produced a temporal walk-off within the medium with the NC geometry. This resulted in the generation of an elliptical harmonic beam profile at the sample surface (see Fig. 1c), which led to the generation of the EUV light sheet at the focal point. The simulated results shown in Fig. 3c and d are reasonably consistent with the experimental results; the relatively small deviation can be attributed to the spherical aberration being omitted in the model.

The results in Fig. 3e and f illustrate the feasibility of high-harmonic wavefront modulation. In the experimental setup shown in Fig. 1a, the distance between the SLM and the lens was set to be 1.35 m. As seen in Fig. 3e, the blazed grating included in the SLM laterally shifted the high-harmonic EUV beams on the millimetre scale. The relationship between the lateral shifting of the EUV beams and the IR diffraction angle was nearly linear in the range of -4 to 4 mrad, thereby proving that the spatial phase distribution can be easily converted via the HHG process. To evaluate the system applicability, we performed dynamic scanning using the EUV beam (Fig. 3f). Over a thousand sequences, the directionality of the EUV beam was confirmed to exhibit exceptional repeatability. Compared with beam control techniques that utilise EUV optics and mechanical components, the proposed scheme ensures high stability because the EUV wavefront is described only by the driving field.

*EUV Bessel beam generation*

HHG on a bare solid surface enables arbitrary manipulation of the high-harmonic beam profile. The experimental scheme adopted to focus the EUV beam entails the generation of a Bessel beam, which is an interferometric pattern that enables tight focusing of coherent light over its diffraction limit. Figure 4a illustrates the experimental scheme to generate converging EUV annular beams. The driving annular beams with different converging angles overlap at the MgO surface to generate spatially-filtered EUV annular beams. The EUV beam profiles were measured by adjusting the microchannel plate (MCP) detector unit in the direction of optical axis. The two annular beams shown in Fig. 4b follow two different parity conservations of the 7th harmonic: $n_1:n_2 =$ 3:4 and 4:3. Figure 4c shows the cross-sectional beam profile of the 7th harmonic in the *x-z* plane. The annular beams propagate with different converging angles ($\theta_b$) to ensure that the resulting peak would be narrower than the single pixel size of our imaging system (80 μm).

To investigate the stability of the system, we measured temporal variation of the beam positioning and yield under a maximum driving intensity of 27 TW/cm$^2$. The EUV photon flux was measured using a copper-beryllium oxide (Cu-BeO) electron multiplier tube (EMT, R5150-10, Hamamatsu Photonics). The photon flux $N$ is calculated as $N = I/(q \times Q \times G)$, where $q$, $Q$, and $G$ denote the electron charge ($1.6 \times 10^{-19}$ C), quantum efficiency (0.17), and gain of the EMT, respectively[7]. Figure 4d shows that a reasonably consistent EUV annular beam profile and position can be maintained over 20 min of operation. Additionally, Fig. 4e shows that the fluctuation in the harmonic yield is insignificant. Because the solid HHG is less sensitive to the phase-matching condition, the environmental noise applied to the sample did not affect the photon flux or beam profile. Owing to the coherent nature of the HHG scheme—as confirmed through diffraction imaging using double slits (**Supplementary Fig. 4**)—and the stability guaranteed by the common-optics configuration, the annular beams could generate an interferometric Bessel beam by near the optical axis.

Generally, the profile of a Bessel beam is experimentally characterised using an optical imaging system or via material processing. Unfortunately, no existing imaging system can obtain an EUV beam profile with nanoscale spatial resolution. Moreover, an EUV beam generated via solid HHG tends to have insufficient peak intensity when applied to process materials. Thus, using the far-field diffraction patterns corresponding to the single slit (see **Methods** for details), we employ a ptychographic iterative engine (PIE) algorithm[26] to reconstruct the amplitude and phase of the EUV field.

Figure 4f shows the intensity profiles of the EUV Bessel beam, which correspond to the spot of the converging annular beam with a converging angle of dashed line of the retrieved 7th harmonic Bessel beam (). Because they were affected by the IR beam block with four bridges, the EUV annular beams formed an interferometric pattern with a sharp peak and four-fold symmetry. This pattern was also confirmed by analysing the 3rd harmonic beam profile, which was directly imaged using the ultraviolet imaging system (**Supplementary Figs. 5 and 6**). The results indicated that a properly designed IR beam block can be used as a spatial Fourier filter that can modify the side lobes of the harmonic beams. The FWHM of the central peak was 390 nm, which is 3.4 times the effective wavelength (115 nm). To the best of our knowledge, such a spot size has not yet been achieved without using EUV optics; furthermore, it is comparable to that achievable by applying focusing techniques using a free-standing zone plate[18].

**Discussion**

While we demonstrated various beam control schemes using a bulk sample, the HHG in a very thin sample can introduce further robustness and controllability with a distinct waveform reproducibility under varying driving intensities and carrier-envelope phase[5]. Specifically, an arbitrarily-shaped driving field hardly introduce distortion of the high-harmonic pulse. Thus, the diffractive-lens function employed in defocused NC-HHG experiments can be applied to further sophisticated applications such as holographic imaging.

In the EUV Bessel beam generation scheme, spectra of the EUV beams (**Supplementary Figure 3**) show that 7th harmonic was predominantly generated in the present geometry. We hardly observed higher-order harmonics as the driving laser polarization is not aligned in the Γ-X direction. Therefore, future investigation may involve materials design considering electronic band structures corresponding to asymmetrically-aligned non-collinear driving fields.

In this study, we employed the ptychographic phase retrieval technique to determine only the electric field of the EUV Bessel beam using a simple single slit sample. However, thanks to the millimetre-scale depth of focus and self-healing nature of the Bessel beam, the proposed technique has the further potential to facilitate multi-dimensional diffraction imaging requiring high photon-flux in a compact area, such as nano-crystallography and holographic tomography. Furthermore, the present scheme can be applied for advanced optical manipulation techniques such as localised optical wave packet generation[27], side lobe suppression[28], and generation of circularly-polarized EUV vortices[29-31], combined with driving field manipulation.

In conclusion, we demonstrated that solid HHG can be effectively utilised for the spatial filtering and the wavefront control of high-harmonic beams without EUV optics. Controllability of the surface-field-driven harmonic beam will be further enhanced by employing a tailored surface. Metasurface- or photon-sieve-based approaches can be effectively utilised to achieve advanced beam control, particularly super-focusing, wavelength-scale holography, and phase-matched HHG along the surface. Furthermore, the effects of the enhanced fields associated with plasmonic nanostructures[7,32] can be investigated for diffractive beam control[33]. We anticipate that the use of surface-field-driven HHG will present a new paradigm for the advancement of EUV and X-ray-based optical systems.

**Methods**

*Non-collinear illumination.* In Fig. 1a, the driving laser pulses are emitted from a Ti:Sapphire regenerated amplifier of a centre 805 nm wavelength with 40 fs pulse duration at a 1 kHz repetition rate (Astrella, Coherent). The spatial light modulator (SLM) is used to steer the propagation direction of the driving laser, colinearly in the first place, by inducing a linearly-varying spatial phase distribution across the laser wavefront. The Wollaston prism splits the linearly polarized driving laser into two orthogonally polarized beams of equal intensity, of which the polarization is controlled by the polariser of Glan type and the half-wave plate as shown in Fig. 1a. Further, the wedged delay line is set to adjust the spatiotemporal overlap of the split driving pulses focused on the MgO crystal through a collimating lens. The beam profiles and spectra of generated high harmonics are measured using an EUV spectrometer & imaging system (see **Supplementary Figure 1**).

*Knife-edge method.* Assuming that the EUV beam has a Gaussian intensity distribution near the focal point, we applied the knife-edge method to measure the beam size in the *x* and *y* directions. By mechanically scanning a sapphire blade across the EUV beam, we used an EUV spectrometer to record the transmitted power as a function of the knife position. The normalised transmitted power distribution in $x$ and $z$ directions was fitted using the error function $P_N(x,z) = \frac{1}{2}\left[1 + \text{erf}\left(\frac{\sqrt{2}x}{w(z)}\right)\right]$, where $w(z)$ is $1/e^2$ beam radius. The cross-sectional Gaussian intensity profile was obtained by differentiating the fitted error function.

*EUV beam propagation simulation.* The simulation of the EUV beam propagation, shown in Figs. 3e and f, was conducted on the basis of the Fourier optics. The electric field at the surface, expressed in Eq. (1), was reduced to its time-independent form using the Helmholtz equation. The experimental data shown in Fig. 2c depict the trend of the effective nonlinear coefficient $q_{eff}$ as a function of the driving intensity. Using the Fresnel transfer function propagator, the electric field after propagation in the $z$ direction can be calculated as follows[34]:

$$E_q(x,y,z+\Delta z) = \mathcal{F}^{-1}\left[\mathcal{F}\{E_q(x,y,z)\}H(f_x,f_y)\right]$$

where the transfer function $H$ is

$$H(f_x,f_y) = e^{jk\Delta z}\exp\left[-j\pi\lambda z(f_x^2 + f_y^2)\right].$$

Here, $k$ and $(f_x, f_y)$ are wavenumber and spatial frequency components, respectively.

***EUV Bessel beam generation experiment.*** In Fig. 4a, the driving IR beam profile was shaped into two concentric annular beams using the annular aperture and axicon. As shown, after being propagated through the focusing lens, the two IR beams become focused and spatiotemporally overlapped in the MgO crystal. The IR beam block, which comprised stainless steel foil supported by three or four bridges, was positioned behind the MgO wafer to selectively extract the EUV beams. In the direct imaging experiment, the results of which are shown in Figs. 4b-d, we used a fused silica axicon with an apex angle $\theta_a = 2.79$ rad (160°) and a lens with a focal length $f$ = 50 mm to obtain a long working distance. In the diffraction imaging experiment using the single slit (Figs. 4f-h), to obtain large converging angles under a millimetre-scale working distance, the axicon ($\theta_a = 2.97$ rad (170°)) and focusing lens ($f$ = 50 mm) were concurrently optimised.

***Ptychographic wavefront retrieval.*** We conducted diffraction imaging experiments using a scanning single slit (see **Supplementary Figure 6**). The slit was fabricated on 15-μm-thick stainless-steel foil using a Yb-fibre femtosecond laser-based micromachining system. The far-field diffraction pattern was recorded as the single slit was shifted in the $x$ and $y$ directions. The slit was vertically aligned in the direction of the shift. The MCP screen was positioned 49–55 mm away from the single slit. This distance is much larger than the applied Fraunhofer distance of 5.6 mm; thus, the far-field condition was satisfied. The single slit was scanned at five points spaced 1.6 μm apart in the $x$ and $y$ directions. Because the length of the slit (3 mm) was significantly larger than the area of interest, it was not necessary to set the scanning coordinates as a grid structure (e.g. 5 × 5). Thus, we only obtained 10 diffraction patterns (i.e. five patterns in each scanning direction) for field retrieval. To increase the spatial resolution, zero padding was implemented outside the diffraction patterns after the noise was removed. The principle of the PIE algorithm is described elsewhere[35]. The update function used in our study was

$$P_{g,n+1}(\mathbf{r}) = P_{g,n}(\mathbf{r}) + \frac{\beta |O(\mathbf{r}-\mathbf{R})|O^*(\mathbf{r}-\mathbf{R})(\psi_{c,n}(\mathbf{r},\mathbf{R}) - \psi_{g,n}(\mathbf{r},\mathbf{R}))}{|O_{max}(\mathbf{r}-\mathbf{R})|(|O(\mathbf{r}-\mathbf{R})|^2 + \alpha)}$$

where $P$, $O$, and $\psi$ are the probe, object, and exit field immediately after the slit, respectively. The slit position $\mathbf{R}$ changes in Cartesian coordinates $\mathbf{r} = (x, y)$. Subscripts $g$ and $c$ indicate whether a field would be guessed or corrected, respectively, and superscript $n$ denotes the iteration number. Considering convergence of the algorithm, tthe parameters $\alpha$ and $\beta$ were optimised as 0.001 and 0.85, respectively. After 200 iterations, the sum squared error was on the order of $10^{-7}$. Additionally, the slit was regarded as a step function without the imaginary part; as such, the values for the transmission area and stainless-steel foil were set to 1 and 0, respectively.

**Acknowledgement**

This work was supported by the National Research Foundation of the Republic of Korea (grant nos. NRF-2012R1A3A1050386, 2020R1A2C2102338, and 2021R1A4A1031660).

**Author contributions**

S.-W. K. planned and oversaw the project in collaboration with Y.-J.K. B.K., S.C., Y. W. K., and S. Won performed the experiments and analysed the data. All authors contributed to the preparation of this manuscript.

**Declaration of competing interest**

The authors declare no competing financial interests.

**Data availability**

The data that support the findings of this study are available from the corresponding author upon reasonable request.

**Figures**

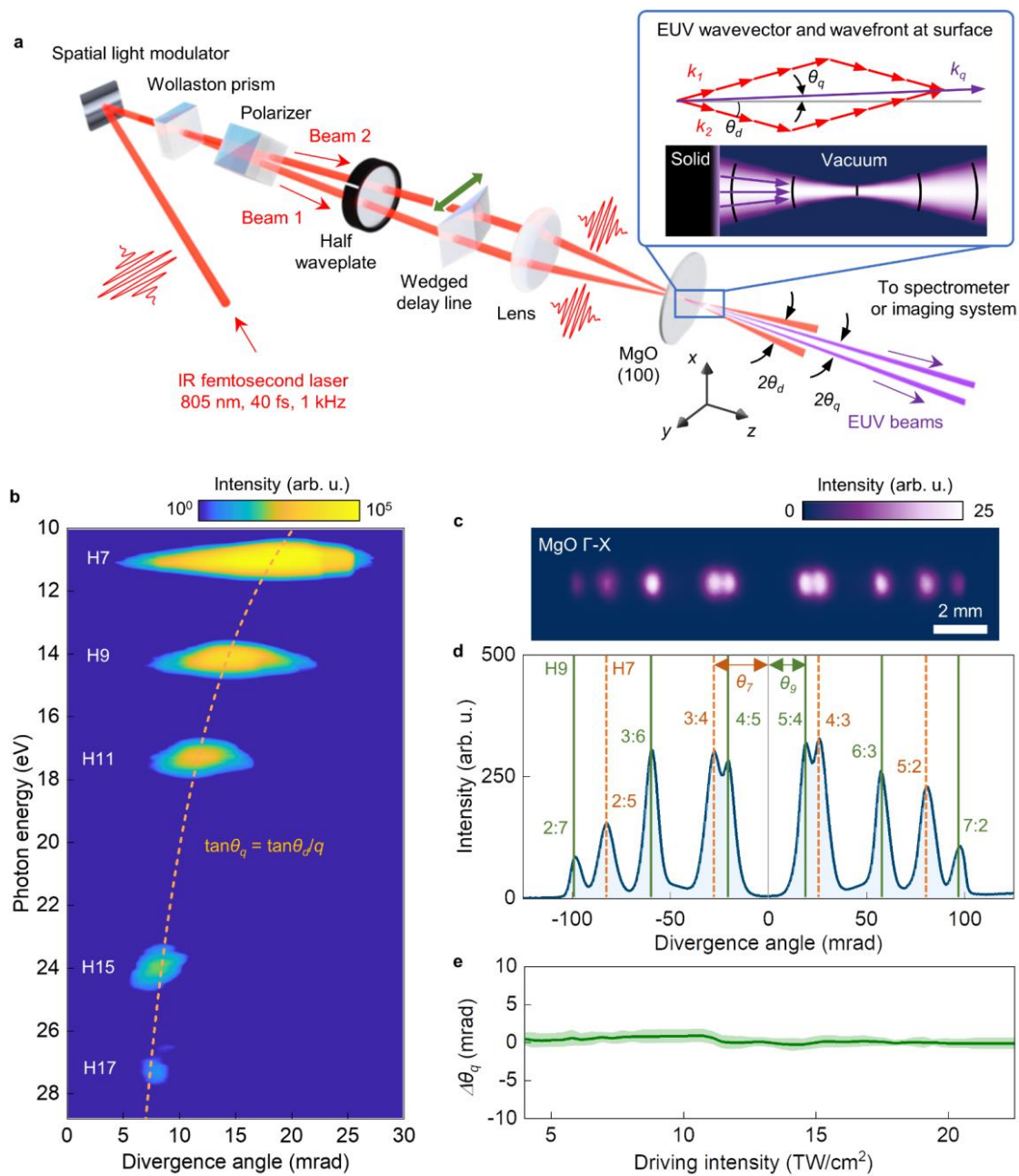

**Figure 1 | Non-collinear high-harmonic generation using an MgO (100) crystal**. **a**, Experimental setup. Inset: high-harmonic wavevector and wavefront. **b**, Divergence angle of 7th, 9th, 11th, and 15th harmonics. The solid orange line indicates that the divergence of the $q$th harmonic satisfies $\tan\theta_q = \tan\theta_d/q$. **c,d**, Far-field harmonic beam profile and cross-sectional intensity distribution, respectively. The orange dashed and green solid lines denote the divergence angles under the parity conservation conditions for H7 and H9, respectively. **e**, Variation in divergence angle $\Delta\theta_q$ of H7–H9 as a function of driving intensity. The light-green area indicates the standard deviation, which was less than 1.3 mrad for all driving intensities. In (**b**)–(**d**), the driving intensity was 23 TW/cm$^2$. The laser polarization was aligned to the Γ-X direction of the MgO (100) crystal.

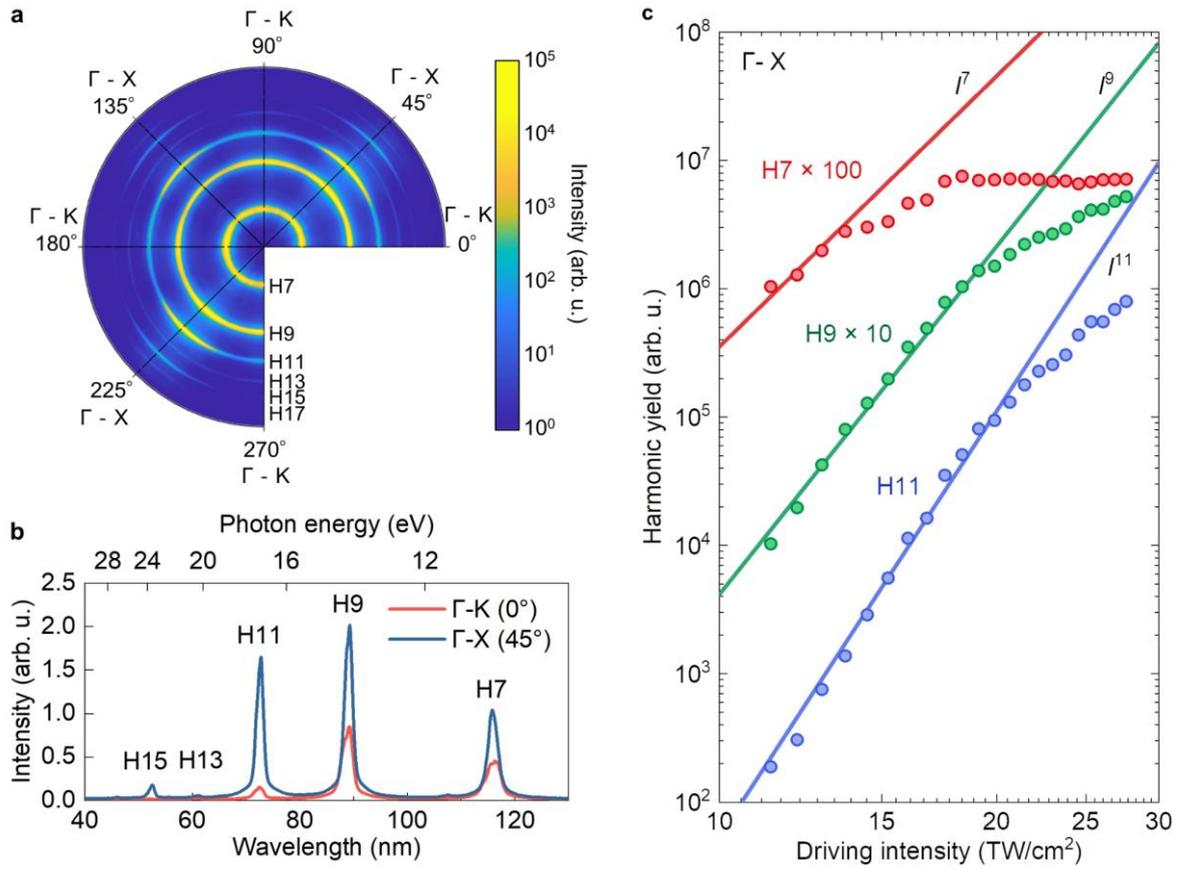

**Figure 2 | High-harmonic spectra showing dependence on crystal orientation and driving intensity in MgO (100) crystal. a**, Harmonic spectrum for various crystal orientations, as investigated via rotating harmonic polarization. **b**, Harmonic spectrum measured under laser polarization aligned in the Γ-K and Γ-X directions. **c**, Harmonic yield as a function of driving intensity. The solid lines correspond to trend curves for $I^q$, where $q$ is the harmonic order.

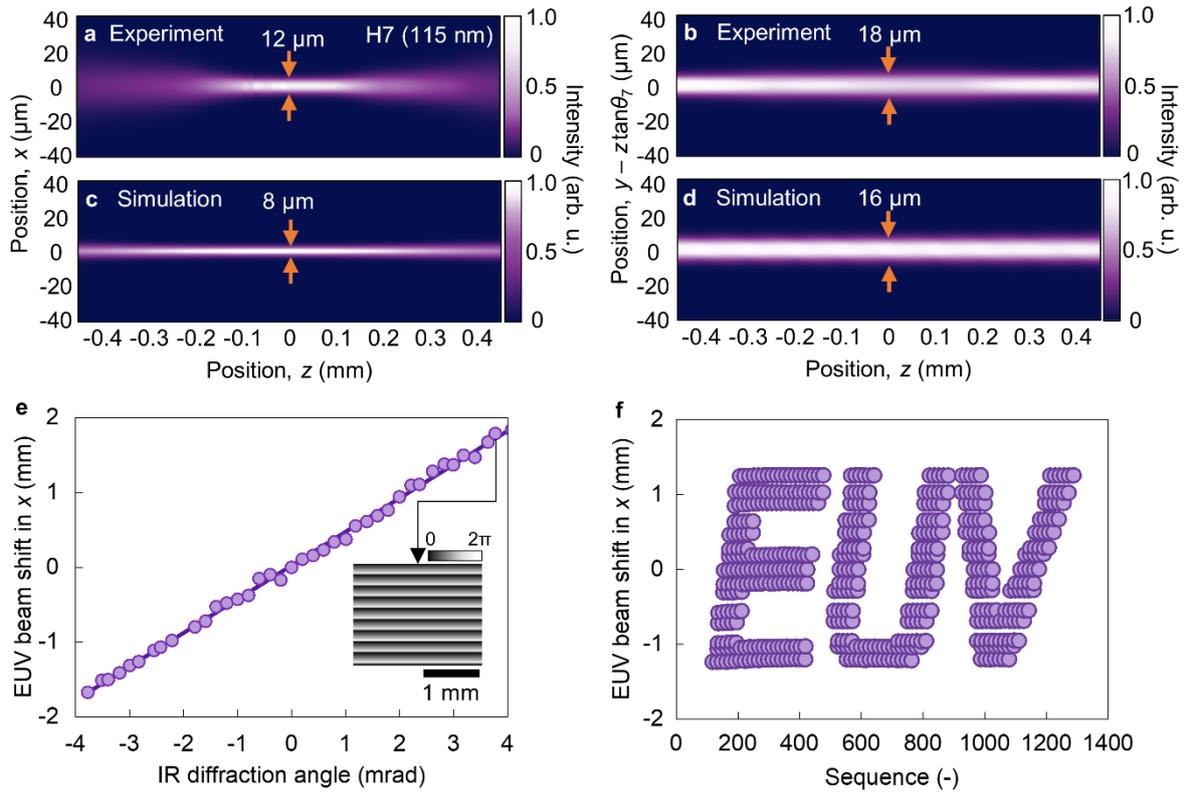

**Figure 3 | Beam profile achieved by focusing the 7th harmonic beam obtained via defocused NC-HHG. a**, **b**, Cross-sectional beam profile obtained in *x* and *y* directions, respectively, using the knife-edge method. **c**, **d**, Simulated results corresponding to (**a**) and (**b**), respectively. Note that a $1/e^2$ beam size is indicated at the foci. In (**b**) and (**d**), the *y* position was shifted by $-z\tan\theta_7$ to correct the harmonic beam divergence. **e**, EUV beam shift as a function of the diffraction angle of the driving IR. Inset: example phase map in the shape of the blazed gratings applied to the SLM. **f**, Dynamic scanning results of the EUV beam in the *x* direction.

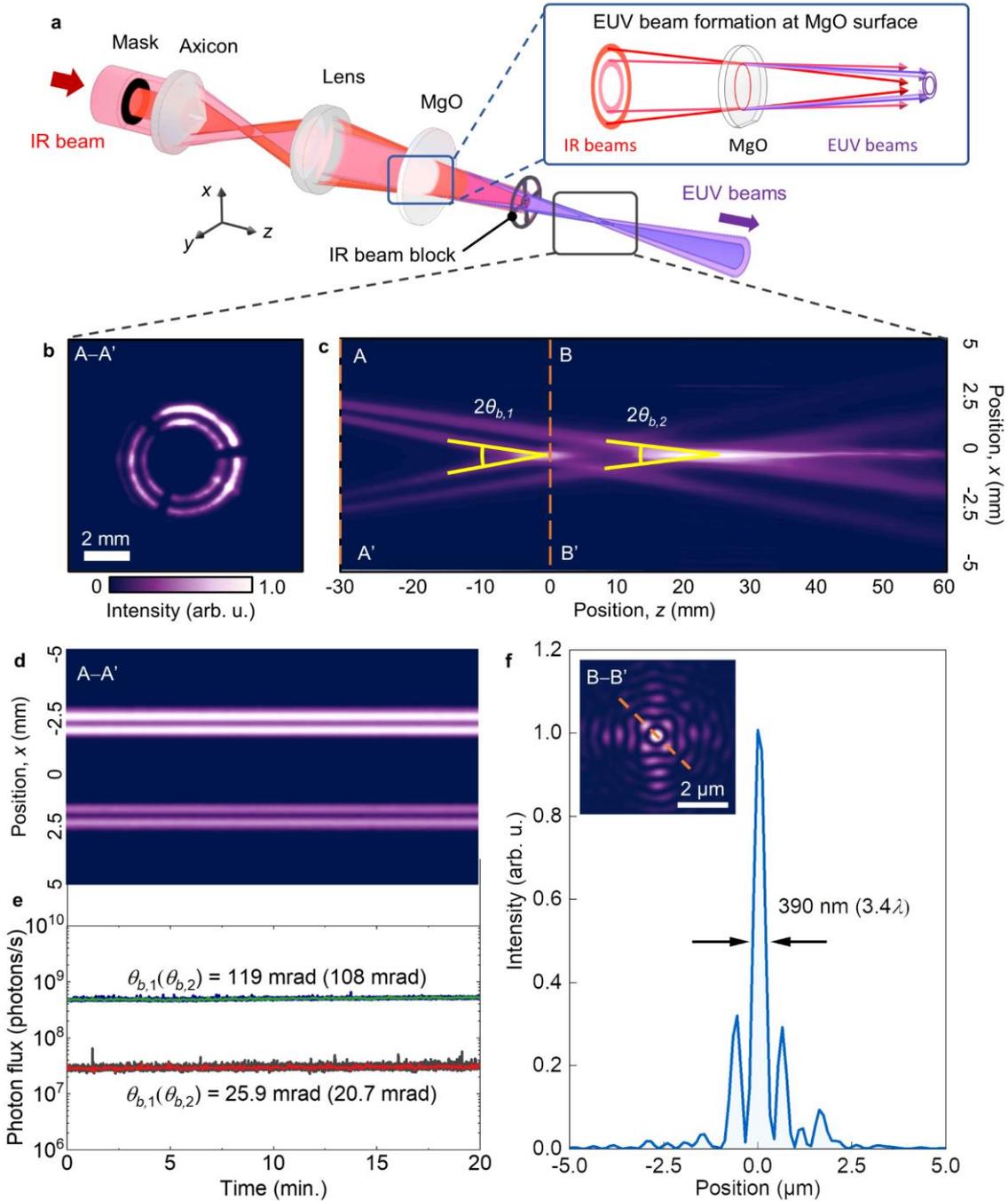

**Figure 4 | EUV Bessel beam generation spatially separated from driving beams. a**, Experimental scheme for EUV Bessel beam generation using two crossing IR annular beams in bulk MgO crystal. **b**, Example of cross-sectional beam profile of H7 beams in x-y plane (A-A' in (**c**)) **c**, Beam profile in *x-z* plane. Two annular EUV beams propagated with different converging angles $\theta_{b,1}$ and $\theta_{b,2}$. **d**, **e**, Variation in cross-sectional beam profile (A-A' in (**c**)), and total harmonic yield over 20 min of operation. Green and red line indicate 100 points averaged values. **f**, Cross-sectional intensity distribution of first focal point (B-B' in (**c**)). Inset: Two-dimensional intensity profiles of EUV Bessel beam at first focal point, as retrieved using ptychographic iterative engine. The converging angle $\theta_{b,1}$ ($\theta_{b,2}$) was 25.9 mrad (20.7 mrad) for (**b**)–(**d**) and 119 mrad (108 mrad) for (**f**).

**Supplementary Information**

**Optics-less beam control of EUV high harmonics generated from solids**

Byunggi Kim[1], Seungman Choi[1], Yong Woo Kim, Seung Jai Won, Young-Jin Kim and Seung-Woo Kim[*]

*Department of Mechanical Engineering, Korea Advanced Institute of Science and Technology (KAIST), 291 Daehak-ro, Yuseong-gu, Daejeon 34141, South Korea.*

*[1]These authors are equally contributed to this work.*

*[*]Corresponding author (swk@kaist.ac.kr)*

## Section 1. Experimental setup for measurement of EUV spectrum and beam profile

Supplementary Fig. 1 shows the schematic illustration of the EUV spectrometer in NC-HHG experiment. We used the Rowland type EUV spectrometer consisting of the toroidal mirror, slits and curved grating. Single-slit (slit 1) with a width of 30 um was placed right before spectrometer system and scanned orthogonally to the optical axis to estimate different divergence angle of each harmonic light. The high-harmonic beam is collected by the toroidal mirror, and passes through a slit (slit 2) at its focal point. The spectrum of the EUV beam is analayzed by the curved grating (groove density: 133.6 mm$^{-1}$, radius of curvature: 998 mm). A combined apparatus consisting of a microchannel plate (MCP, XUV-2040, BrightView) and the charge coupled device camera (DH420A-FO-195, Andor) detected the spread light as a form of line spectrum, providing a spectral resolution of 0.16 nm. For measurement of the EUV beam profiles in the manuscript, we directly imaged EUV beams by using a MCP (BOS-40-6, Beam Imaging Solution) without the spectrometer. The whole system is installed under the high vacuum environment below 10$^{-5}$ Torr.

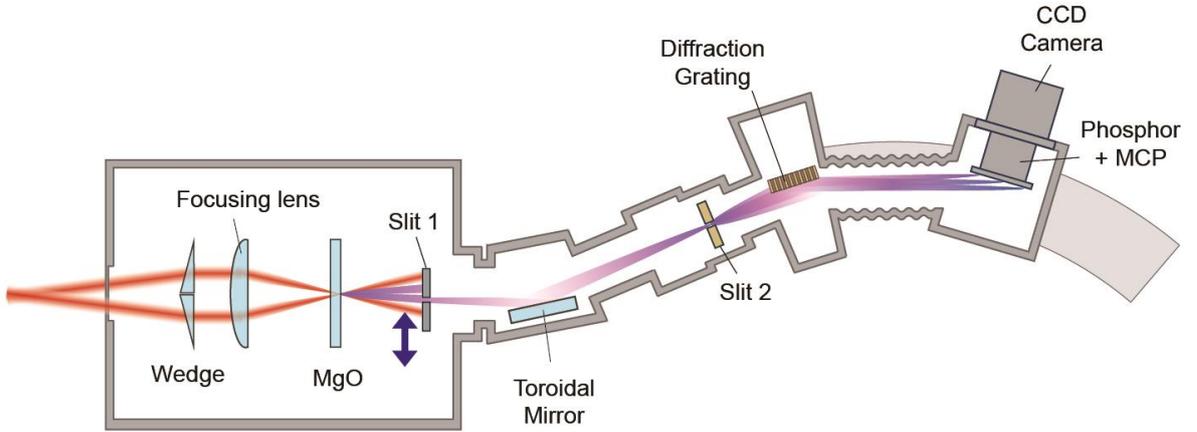

**Supplementary Figure 1.** EUV spectrometer setup for measuring angular dispersion in NC-HHG. The slit 1 with 30-um width was scanned to characterize dispersion angle of the harmonic beams. The high-harmonic beam after the slit 1 is analyzed by Rowland type spectrometer.

## Section 2. Estimation of driving field intensity

We obtained the driving pulse intensity by measuring pulse energy and beam profile. The beam profile at the sample was measured by a CCD camera. Assuming the temporally Gaussian pulse, the spatial intensity distribution of the driving pulse $I(x,y)$ is calculated on the basis of the following relationship between pulse energy $\epsilon_p$:

$$\epsilon_p = \sum_{x,y} I(x,y) \cdot \tau_p / 0.94 \tag{S1}$$

where $\tau_p$ denotes FWHM pulse duration, which is 40 fs in the present context.

### Section 3. NC-HHG from *α*-Quartz crystal

We observed high-harmonic beams and spectrum generated from the Γ-M direction of the *α*-Quartz crystal. Supplementary Figure 2a shows that even order harmonics are produced due to lack of the inversion symmetry. It attributes to centered-projection of the EUV beam, when the number of photons absorbed from each driving pulse is same, given as $q_1 = q_2$. It is shown as the zero-divergence beam in Supplementary Fig. 2b. As it includes all the even-order harmonics components, the broadband EUV spectrum can be spatially separated from the driving beams directly after the crystal sample.

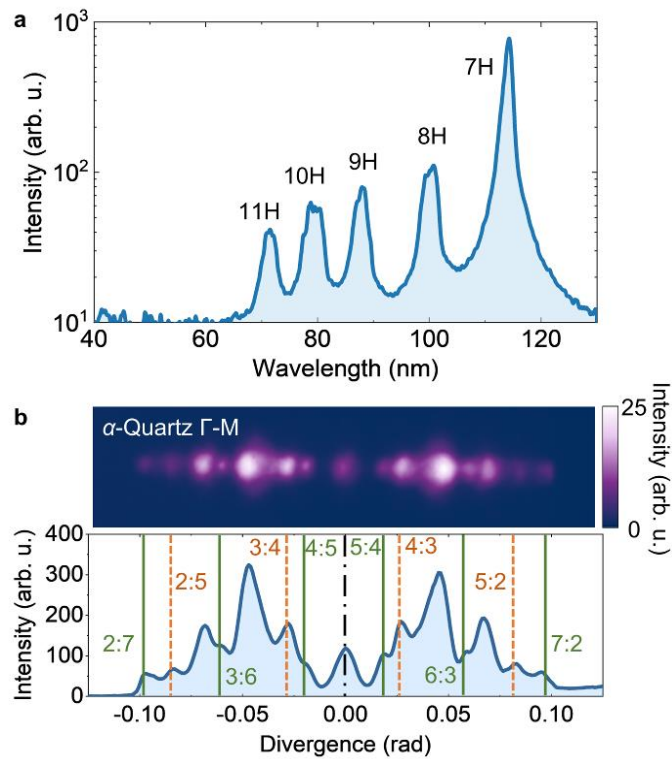

**Supplementary Figure 2.** Non-collinear high harmonic generation from *α*-Quartz crystal (Γ-M). **a**, high-harmonic spectrum. **b**, Spatially separated EUV beam profiles. Dashed orange line and solid green line indicate 7th and 9th harmonic beams, respectively. Black dotted dashed line is zero divergence angle where even-order harmonics propagate.

## Section 4. Spectral response and spatial coherence of the high-harmonic EUV beams generated by annular driving beams

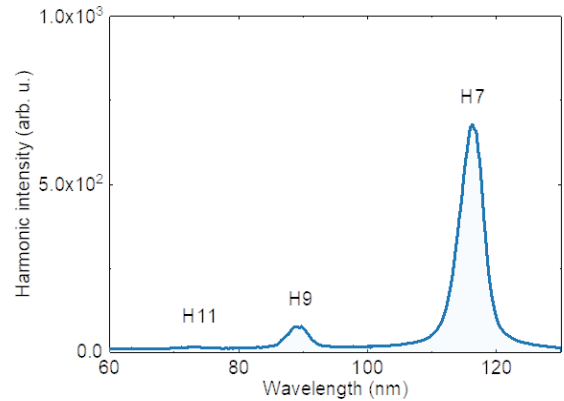

**Supplementary Figure 3.** Spectral response of the high-harmonic EUV beam via annular driving beams

Spatial coherence of the 7th harmonic EUV beam generated by the annular-shaped driving beams was investigated based on the far-field diffraction from the double slit with the experimental setup shown in Fig. 4a of the main text. Supplementary Figure 4a shows a scanning electron microscope (SEM) image of the double slit with 18~20 μm-width each and 50 μm-interval. In Supplementary Fig. 4b, the far-field diffraction pattern from the double slit has clear interference fringe which demonstrates spatial coherence of the generated EUV beams. The fringe width was measured as 130 μm, corresponding well with Young's double slit calculation given by $\frac{\lambda L}{d}$. Here, $\lambda$, $L$, and $d$ are wavelength, distance from slit to screen, and slit interval, respectively.

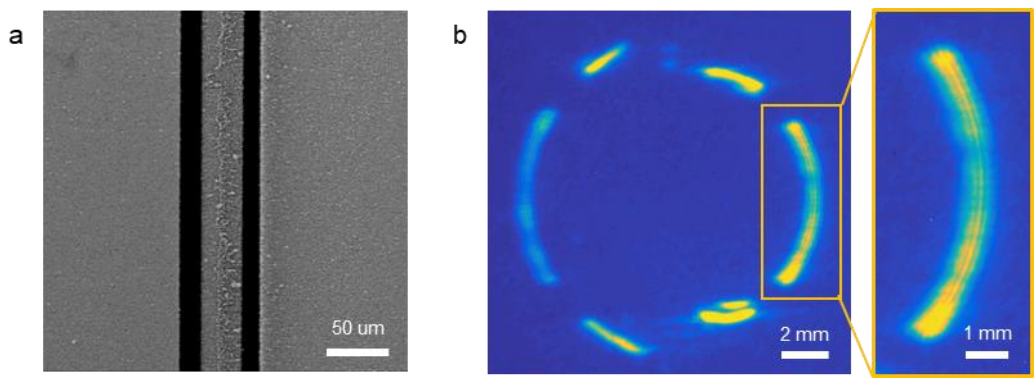

**Supplementary Figure 4.** Interference of the EUV conical beam through a double slit. a, The vertical double slit fabricated by Ytterbium-doped fiber femtosecond laser (Lasernics, FUPL-250-6). b, Far-field diffraction pattern through the slit. The measured fringe patterns demonstrated spatial coherence of the generated EUV conical beam.

**Section 5. Comparison between Bessel beam of THG**

Low-order harmonics can be characterized by refractive optics and CCD camera with good sensitivity under non-vacuum environment. Therefore, it can be used as a good reference for estimating high-harmonic beam profile. To compare with the phase retrieval results based on the ptychographic iterative engine (PIE) (Fig. 4 in the main text), we measured the third-harmonic (268 nm) beam profiles using the ultraviolet imaging system illustrated in Supplementary Figure 5a. The combined setting with the UV objective lens (50× M Plan UV, Mitutoyo) and an UV CCD camera (JAI, CM-140MCL-UV) was used to refocus and get a magnified-image of the third-harmonic beam. To avoid possible illumination of diffracted IR beam, a bandpass filter (BPF) with a center wavelength of 266 nm and FWHM band width of 10 nm was placed before the UV lens. Supplementary Fig. 5b shows annular beams with different parity conservation conditions ($q_1$:$q_2$ = 1:2 and 2:1) form interferometric patterns at different axial position $z$. The magnified beam profiles at the foci shown in Supplementary Fig. 5c and d agree well with the 7th-harmonic Bessel beam profile in Fig. 4h of the main text. The third-harmonic beam profiles also have the 4-fold symmetry affected by the IR beam block. Therefore, we could consider that the PIE successfully retrieved EUV beam profile based on the experimental method using a micro slit. Supplementary Figure 6 shows experimental setup for diffraction imaging and extended results of the EUV field retrieval on the various axial positions. The EUV Bessel beam has quasi-concentric phase distribution and two focal points on the optical axis between the foci of the third-harmonic beams.

We also conducted photolithography using the third-harmonic Bessel beam. Positive photoresist (ZEP520A) was spin-coated on silicon-wafer at 4000 rpm for 60 sec so as to get 300-nm thickness. The resist was pre-baked and post-baked at 140℃ for 3 min, respectively. Supplementary Figures 5e and f indicates that the surface of the photoresist has clear patterns of the Bessel beam profile after being illuminated by the third-harmonic beam for 5 min. We believe that by improving conversion efficiency of the high-harmonics and intensity ratio between the main peak and side lobes, the laboratory-scale EUV lithography with a sub-100 nm resolution will be enabled on the basis of the solid HHG.

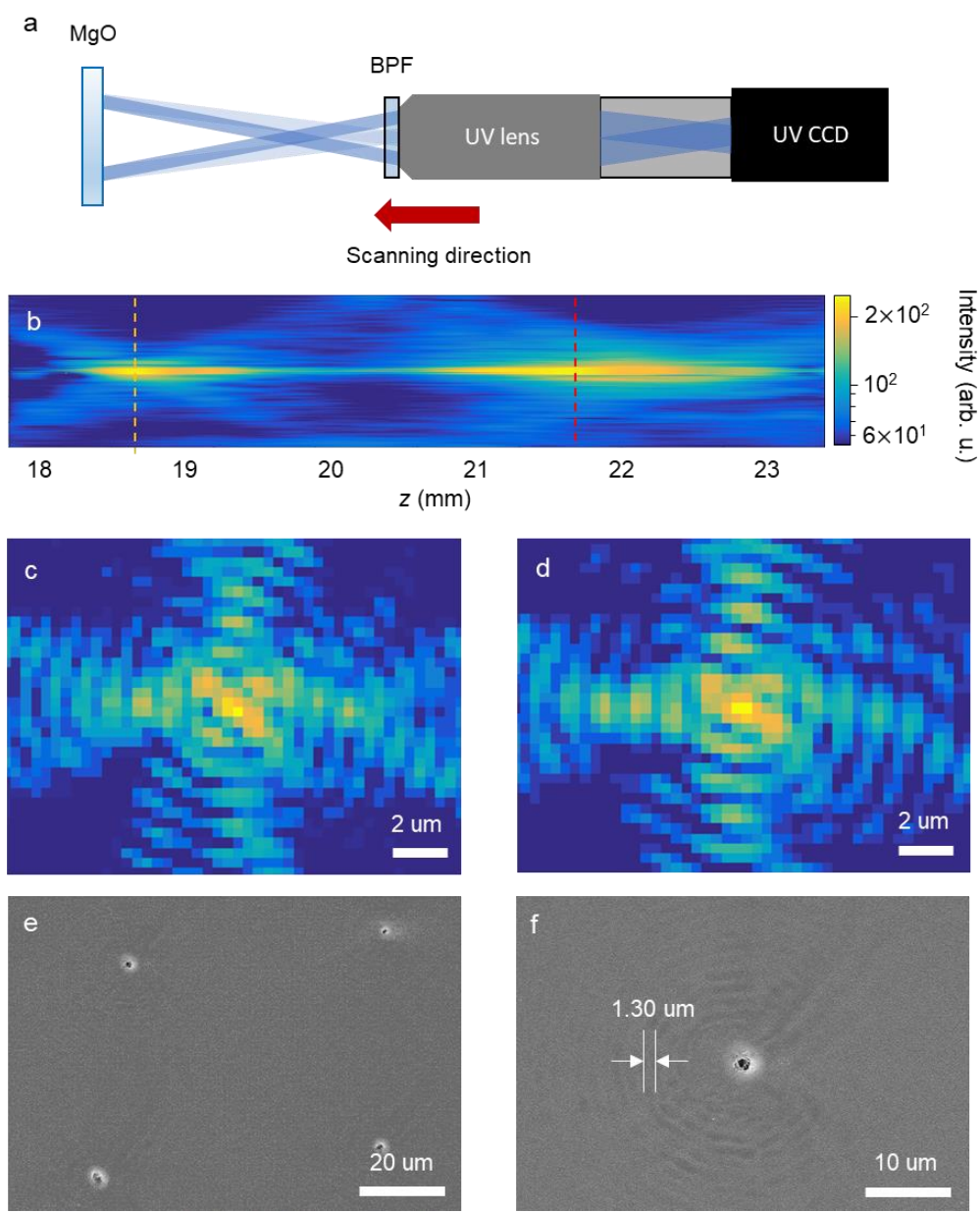

**Supplementary Figure 5.** Third-harmonic Bessel beam characterization. **a**, Schematic illustration of ultraviolet imaging system. **b**, Direct imaging of 3rd harmonic (267 nm) conical beam along propagation direction. $z$ is distance from the MgO sample. **c**, **d**, the third-harmonic Bessel beam profile at forward and backward focus (which is represented by orange and red dashed lines in (**b**)), respectively. **e**, **f**, SEM images of the third-harmonic lithography. Interval of the side-lobes engraved in photoresist was used for scale calibration of the third-harmonic beams imaged in (**b**)-(**d**).

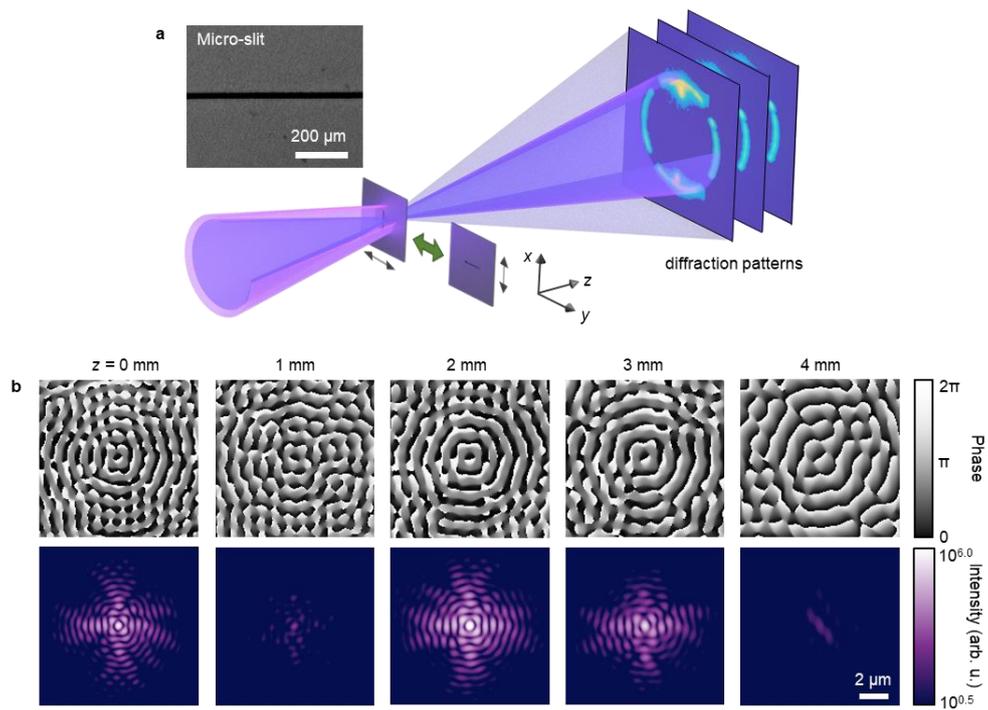

**Supplementary Figure 6.** Ptychographic EUV beam reconstruction **a**, Experimental setup for EUV Bessel beam retrieval. **b**, Two-dimensional phase and intensity profiles of EUV Bessel beam along propagation direction *z*. The upper and lower images show the retrieved phases and intensities at the *z*-position indicated on the top, respectively.